\documentclass{article}
\usepackage{mrs2005,epsfig,amsmath,amssymb,natbib}

\newcommand{\ie}{\textit{ie.}}

\newcommand{\HI}{H{\footnotesize I}}
\newcommand{\Htwo}{H$_2$}
\newcommand{\subHI}{_{\text{H{\tiny I}}}}

\newcommand{\Mtwo}{M_{200}}
\newcommand{\Rmin}{{\cal R}_{\text{min}}}
\renewcommand{\sun}{$_{\odot}$}

\begin{document} 
\title{THE HI MASS--STAR FORMATION RATE RELATION
\\ AND SELF-REGULATED STAR FORMATION IN DWARF GALAXIES}

\author{Edward N Taylor}
\affil{Leiden Observatory} 

\begin{abstract} 
We have developed a simple, static model designed to place a very
solid lower limit on the star formation rate (SFR) expected in a dwarf
disk galaxy, which leads to the prediction of a previously
undocumented relation between \HI\ mass, $M\subHI$, and SFR.  Over the
mass range $10^8$---$10^{10}$ M\sun , a wide variety of galaxies are
observed to follow such a relation --- SFR $\propto M\subHI^{1.4}$ ---
with the same slope and similar scatter to our prediction.  Within the
model, this relation is a manifestation of self-regulating star
formation (SF), in which the ISM is kept warm and stable by a UV
interstellar radiation field (ISRF) that is maintained by constant
regeneration of O---B stars.  Regardless of the actual mode of star
formation, it seems that the majority of dwarfs are presently forming
stars in the same way.
\end{abstract} 
 
\section{Introduction} 
 
The work presented is derived from an attempt to understand a
surprising result from the \HI\ Parkes All-Sky Survey (\HI PASS): the
non-detection of isolated, galactic-scale concentrations of \HI\ with
no significant stellar counterpart --- baryonic dark galaxies or
`galaxies without stars' \citep{DoyleEtAl}.
We have presented this study in \citet{TaylorWebster}: we find that an
\HI\ surface density as low as several M\sun\ pc$^{-2}$ is sufficient
to self-shield against \Htwo\ photodissociation by the cosmic
background radiation; we argue that within this self-shielded region,
a relative \Htwo\ abundance of as little as 10$^{-4}$ is sufficient to
initiate a transition to a cold ($T \lesssim 300$ K) phase, which
greatly increases the chances of gravothermal (Toomre) instability.
In short, we conclude that dark galaxies are, at best, exceedingly
rare, because {\em there is no mechanism to prevent star formation};
wherever possible, self-shielding \HI\ will develop an \Htwo\
component that is sufficient to initiate gravothermal instability.

The situation changes once SF is initiated: short-lived O--B stars
then provide a second source of photodissociating radiation, much as
in photodissociation regions in our own galaxy.  Using this idea of
self-regulated star formation --- the gravothermal stability of an
\HI\ cloud preserved by photodissociative frustration of \Htwo\
cooling \citep{LinMurray, OmukaiNishi} --- we then set out to
determine the minimum SFR in a galactic-scale concentration of \HI .

The details of this calculation are presented in \citet{TaylorWebster}
and we do not repeat them here.  In essence, the argument is this: we
solve for the UV (ISRF) energy density required prevent \Htwo\
production/cooling and subsequent instabilities; knowing the UV output
and lifetime of a massive star, and assuming a standard IMF, it is
then possible to determine the SFR required to provide this UV
radiation, $\Rmin$.  We emphasise that since we were interested
primarily in `galaxies without stars' (\ie\ the complete preventation
of star formation) we deliberately tailored our assumptions to produce
a lower limit on the predicted \Htwo\ abundance and rate of cooling,
so making star formation as difficult as possible: in particular, we
have allowed `maximum damage' by photodissociating photons by ignoring
internal extinction, and ignored all effects of dust and metals.

\section{The $M\subHI$--SFR Relation}

We model the gas distribution using the formalism of
\citet{MoMaoWhite}, in which each cloud is described by three
parameters: the virial mass, $\Mtwo$, the spin parameter $\lambda$,
and the disk mass fraction $m_d$; in all cases, $M\subHI / m_d \Mtwo$
is between 0.56 and 0.70.  For each combination of $\lambda$ and
$m_d$, $\Rmin \propto M\subHI^{1.4}$: our predicted $M\subHI$--SFR
relation is shown in Figure 1, along with observed values taken from
the literature.  We note with interest that this relation has the same
power as the empirical Schmidt SF law, SFR$/dA \propto
\Sigma\subHI^{1.4}$.

The slope of the relation is driven by the distribution of the densest
gas in galaxy centres, since it is this gas that requires the most
radiation to maintain thermal stability.  The scatter is set primarily
by the distribution of galactic spins, as well as variations in the
relative mass of the disk in comparison to the halo, since, for a
fixed mass, these determine the density near the centre.  Because the
model was designed to place a lower limit on the SFR, it cannot
predict the zero-point of such a relation.  Within the self-shielded
region, the \Htwo\ cooling rate is typically less than 1---10 percent
of the net cooling rate; a significantly higher ISRF might be required
to balance these other cooling mechanisms.  Moreover, if there were,
for example, a systematic run with mass in the relative contribution
of metal-line cooling in comparison to \Htwo\ cooling, the inclusion
of metals might alter the predicted slope of the relation.  We stress
that this would only be true, however, if metal-line cooling were
important in {\em initiating} the transition from warm to cold.

\section{Self-Regulated Star Formation in Dwarf Galaxies?}

\begin{figure}  
\begin{center}
\epsfig{figure=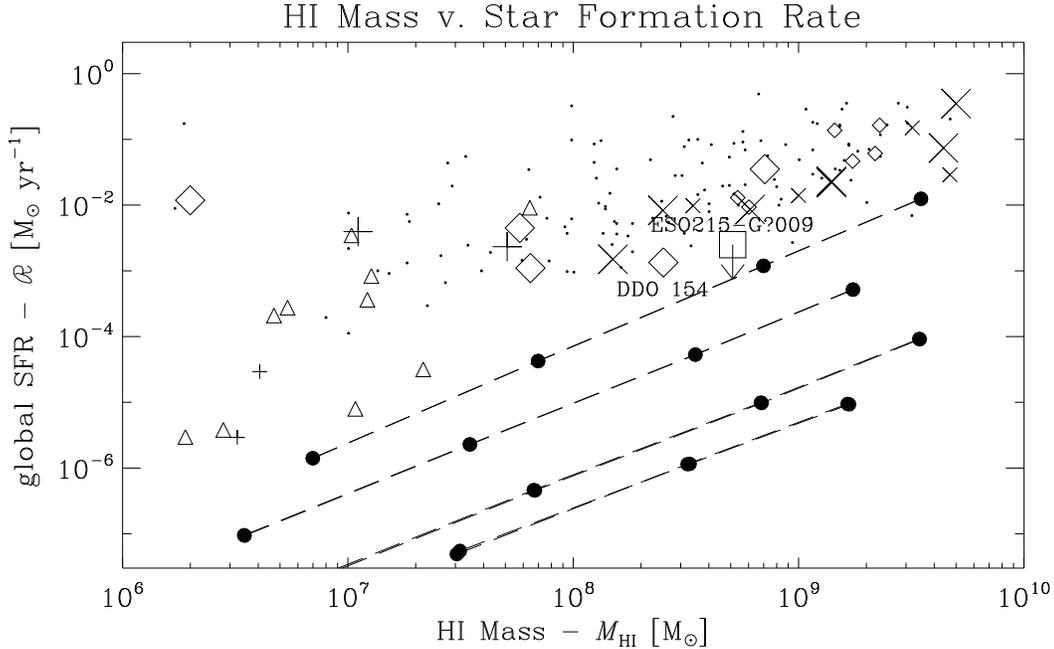,width=14.cm}  
\end{center}
\caption{\footnotesize{Minimum SFR required for stability, $\Rmin$,
    plotted against \HI\ mass, $M\subHI$.  Like models are connected
    with a dashed line to guide the eye: from top to bottom, models
    are ($\lambda, ~m_d$) = (0.04, 0.10), (0.04, 0.05), (0.10, 0.10),
    (0.10, 0.05).  Note that the top and bottom pairs of lines
    represent the 50 and 90 percent points of the expected spin
    distribution, respectively, and that we have deliberately
    attempted to place a firm lower bound on $\Rmin$.  For comparison,
    literature values of $M\subHI$ and SFR for a variety of dwarf
    galaxies are overplotted --- {\em plusses}: 13 Im and Sm dwarfs
    from \citet{HunterElmegreenBaker}; {\em crosses}: four \HI- bright
    dwarfs from \citet{YoungEtAl}; {\em diamonds}: seven LSB and four
    ``normal'' dwarfs from \citet{vanZeeEtAl}; {\em points}: 121
    members of a sample of irregular galaxies (points), spanning more
    than 8 mag in absolute magnitude and surface brightness, from
    \citet{HunterElmegreen}; {\em triangles}: 10 faint dwarf galaxies
    from \citet{BegumEtAl}; {\em square}: the extremely \HI -rich dwarf
    ESO215-G?009 (large square) \citep{WarrenJerjenKoribalski}.}
} \label{fig:MHI-SFR}\end{figure}

Even though the observational samples were constructed to include
widely disparate populations, these populations are not readily
distinguished in Figure \ref{fig:MHI-SFR} for \mbox{$M\subHI \gtrsim
10^8$ M\sun,} nor are `extreme' objects like DDO 154 and ESO215-G?009
distinguished from `normal' dwarfs in the $M\subHI$---SFR plane.  In
the mass range $M\subHI \sim 10^8$---$10^{10} M_{\odot}$, it seems that
essentially all (\HI\ detected) galaxies are presently forming stars
in the same way: galaxies follow a tight sequence in the
$M\subHI$--SFR plane.  Moreover, we argue that the general agreement
between the character of the predicted and observed $M\subHI$---SFR
relations lends weight to the physical picture on which our simple
model is based: the results in Figure \ref{fig:MHI-SFR} suggest that
the well observed population of dwarf galaxies represent the minimum
rate of ISRF-regulated SF in galaxies.

\vfill 
\end{document}